\documentclass[twocolumn,prb,showpacs,superscriptaddress,preprintnumbers,amsmath,amssymb,floatfix]{revtex4}
\usepackage{graphicx}
\usepackage{dcolumn}
\usepackage{bm}
\usepackage{color}

\begin{document}

\title{ Intricacies of the Co$^{3+}$ spin state in Sr$_2$Co$_{0.5}$Ir$_{0.5}$O$_4$: an x-ray absorption and magnetic circular dichroism study}
\author{S.~Agrestini}
  \affiliation{Max Planck Institute for Chemical Physics of Solids,
     N\"othnitzerstr. 40, 01187 Dresden, Germany}
\author{C.-Y.~Kuo}
  \affiliation{Max Planck Institute for Chemical Physics of Solids,
     N\"othnitzerstr. 40, 01187 Dresden, Germany}
\author{D.~Mikhailova}
  \altaffiliation[Present address: ]{Karlsruhe Institute of Technology (KIT), Institute for Applied Materials (IAM), Hermann-von- Helmholtz-Platz 1, D-76344 Eggenstein-Leopoldshafen, Germany}
 \affiliation{Max Planck Institute for Chemical Physics of Solids,
     N\"othnitzerstr. 40, 01187 Dresden, Germany}
  \affiliation{Institute for Complex Materials, IFW Dresden,
     Helmholtzstr. 20, D-01069 Dresden, Germany}
\author{K.~Chen}
  \affiliation{Synchrotron SOLEIL, L'Orme des Merisiers, Saint-Aubin, 91192 Gif-sur-Yvette, France}
\author{P. Ohresser}
  \affiliation{Synchrotron SOLEIL, L'Orme des Merisiers, Saint-Aubin, 91192 Gif-sur-Yvette, France}
\author{T.W.~Pi}
  \affiliation{National Synchrotron Radiation Research Center (NSRRC), 101 Hsin-Ann Road, Hsinchu 30077, Taiwan}
\author{H.~Guo}
  \affiliation{Max Planck Institute for Chemical Physics of Solids,
     N\"othnitzerstr. 40, 01187 Dresden, Germany}
\author{A.~C.~Komarek}
  \affiliation{Max Planck Institute for Chemical Physics of Solids,
     N\"othnitzerstr. 40, 01187 Dresden, Germany}
\author{A.~Tanaka}
  \affiliation{Department of Quantum Matter, ADSM, Hiroshima University, Higashi-Hiroshima 739-8530, Japan}
\author{Z.~Hu}
  \affiliation{Max Planck Institute for Chemical Physics of Solids,
     N\"othnitzerstr. 40, 01187 Dresden, Germany}
\author{L.~H.~Tjeng}
  \affiliation{Max Planck Institute for Chemical Physics of Solids,
     N\"othnitzerstr. 40, 01187 Dresden, Germany}

\date{\today}
\begin{abstract}
We report on a combined soft x-ray absorption and magnetic circular dichroism (XMCD) study at the Co-$L_{3,2}$ on the hybrid 3$d$/5$d$ solid state oxide Sr$_2$Co$_{0.5}$Ir$_{0.5}$O$_4$ with the K$_2$NiF$_4$ structure. Our data indicate unambiguously a pure high spin state $(S=2)$ for the Co$^{3+}$ (3$d^6$) ions with a significant unquenched orbital moment $L_z/2S_z=0.25$ despite the sizeable elongation of the CoO$_6$ octahedra. Using quantitative model calculations based on parameters consistent with our spectra, we have investigated the stability of this high spin state with respect to the competing low spin and intermediate spin states.

\end{abstract}

\pacs{71.70.Ch, 75.47.Lx, 78.70.Dm, 72.80.Ga}

\maketitle

Cobalt compounds have aroused a great deal of attention in the scientific community because of the complex and large diversity of physical phenomena displayed, including metal-insulating transitions ~\cite{raccah67a,Martin97,Imada98}, superconductivity ~\cite{takada03a}, large magneto-resistance ~\cite{perez97a} and high thermoelectric power ~\cite{terasaki97a}. This richness of electronic and magnetic properties is closely related not only to the possibility of stabilizing cobalt in different valence states but also to its ability to present different spin states, the so-called spin-state degree of freedom ~\cite{Sugano,Goodenough71}. For example, in an octahedral coordination, Co$^{3+}$ ions, which have the $d^6$ configuration, can exist in three possible spin states: a high spin (HS) state ($S=2$, $t_{2g}^4e_g^2$), a low spin (LS) state ($S = 0$, $t_{2g}^6e_g^0$) and even an intermediate spin (IS) state ($S = 1$, $t_{2g}^5e_g^1$) ~\cite{Sugano,Goodenough71b,Haverkort06}.

This spin state degree of freedom is evident in LaCoO$_3$ where the Co$^{3+}$ ions have a non-magnetic LS ground state and undergo a gradual transition with increasing temperature to a magnetic spin state ~\cite{Heikes64,Blasse65,Naiman65}. The nature of the magnetic spin state (IS or HS) was heavily disputed in literature for over four decades, till an XMCD study clearly demonstrated it to be HS~\cite{Haverkort06}. Calculations of the Co$^{3+}$ energy level diagram show that the LS (HS) state can be stabilized by a large (small) value of the crystal field 10$Dq$. The IS is always higher in energy for CoO$_6$ octahedra close to regular, like in LaCoO$_3$ ~\cite{Haverkort06}.

However, the IS state, with one electron in the $e_g$ states, is Jahn-Teller active and, hence, can gain energy and become the ground state in the presence of a sufficiently large distortion of the local structure~ \cite{Maris03}. For this reason the spin state of the Co$^{3+}$ ions in layered cobaltates, where the elongated distortion of the CoO$_6$ octahedra favors and may stabilize the IS state, have been subject of intense debate. In the case of layered La$_{2-x}$Sr$_x$CoO$_4$, contradicting scenarios for the Co$^{3+}$ ions were considered to interpret the complex structural, magnetic and transport properties of the system as a function of Sr doping: LS Co$^{3+}$, IS Co$^{3+}$~ \cite{Zaliznyak00,Zaliznyak01,Chichev06}, HS-to-IS transition \cite{Moritomo97}, and mixing of HS/IS ~\cite{Horigane07,Horigane08}. Only recently the spin state of Co$^{3+}$ in layered La$_{2-x}$Sr$_x$CoO$_4$ was demonstrated by X-ray absorption spectroscopy (XAS) studies to be LS ~\cite{Chang09,Merz11} for x~=~0.5 and a mixture of LS/HS for x~$\geq1$ \cite{Merz11,Guo16,Li16}. Band formation has been proposed to possibly provide another route for the stabilization of IS Co$^{3+}$ ~\cite{Korotin96,Ou16}.

In this context, the recently reported synthesis of the layered Sr$_2$Co$_{0.5}$Ir$_{0.5}$O$_4$, where the cobalt ions have been suggested to have the 3+ valence, is very interesting~\cite{Mikhailova17}. In fact, on one hand, the introduction of Co$^{3+}$ ions in a compound with relatively large lattice parameters, like Sr$_2$IrO$_4$, should lead to a reduced crystal field making the LS state energetically less favourable with respect to the HS state. On the other hand, the elongation of the CoO$_{6}$ octahedra, revealed by Co-$K$ EXAFS \cite{Mikhailova17}, should favor a Jahn-Teller active IS state. Moreover, the hybridization of the Co 3$d$ orbitals with the spatially extended Ir 5$d$ orbitals could enhance the importance of band formation. Hence, the layered Sr$_2$Co$_{0.5}$Ir$_{0.5}$O$_4$ provides the opportunity to investigate whether the combined effect of CoO$_6$ elongation and band formation can stabilize the IS state as the ground state.
Experimentally, Sr$_2$Co$_{0.5}$Ir$_{0.5}$O$_4$ is an insulator with antiferromagnetic interactions with an effective moment of $\mu_{eff} = 3.2(1)~ \mu_B$/f.u.~\cite{Mikhailova17}. Similar to the parent compound Sr$_2$IrO$_4$ \cite{Kim2008,Kim2009}, also in Sr$_2$Co$_{0.5}$Ir$_{0.5}$O$_4$ the large spin-orbit coupling is expected to be the leading energy scale in determining the ground state of the Ir$^{5+}$ ions. If these expectations are correct, then the Ir$^{5+}$ ions should have a singlet $J=0$ non-magnetic ground state and provide only a Van Vleck contribution to the magnetism of the system. The measured value of $\mu_{eff}$ is lower than the value $3.46 ~ \mu_B$/f.u. expected for a spin-only HS Co$^{3+}$ ($3.6 ~ \mu_B$/f.u., if the orbital moment $L=1$ is also considered) and significantly larger than the value $2.00~ \mu_B$/f.u. expected for a spin-only IS Co$^{3+}$. Co-$K$ XAS of Sr$_2$Co$_{0.5}$Ir$_{0.5}$O$_4$ indicated a spin state of the Co$^{3+}$ ion higher than LS ~\cite{Mikhailova17}. Yet, as pointed out by Vanko et al. \cite{Vanko06}, the Co-$K$ XAS cannot distinguish unequivocally between IS and HS scenarios due to the relatively small differences in the calculated lineshapes of the pre-edge features. Hence, on the base of the experimental data available in literature both scenarios of a pure HS or a mixture IS/HS are equally possible for the Co$^{3+}$ ions in Sr$_2$Co$_{0.5}$Ir$_{0.5}$O$_4$. The fact that Sr$_2$Co$_{0.5}$Ir$_{0.5}$O$_4$ was synthesized as a single phase and stoichiometric without oxygen deficiency is also important, as oxygen vacancies, which are often present in other cobalt oxides, can cause the formation of CoO$_5$ pyramids (or CoO$_4$ tetrahedra as in Sr$_2$Co$_{1.2}$Ga$_{0.8}$O$_5$ ~\cite{Istomin15}) which may stabilize the HS state~\cite{Hu04,Belik06} or provide the space for stabilizing the HS in a neighbour octahedral Co$^{3+}$ ion ~\cite{Li11,Hu12,Chen14,Istomin15}.

Here we report on an investigation of the local magnetism of the Co$^{3+}$ ions in Sr$_2$Co$_{0.5}$Ir$_{0.5}$O$_4$ by employing soft XAS and XMCD spectroscopies, two powerful techniques that are element selective and, through the lineshapes of the spectra, extremely sensitive to the valence, spin, and orbital or crystal field state of the ion under study. Furthermore the application of sum rules to the XMCD data allowed us to have quantitative information on the orbital moment with respect to the spin moment on the selected ion.

Synthesis of the layered Sr$_2$Co$_{0.5}$Ir$_{0.5}$O$_4$ was carried out from stoichiometric powder mixtures of home-made Co$_3$O$_4$ with IrO$_2$ (Umicore) and SrCO$_3$ (Alfa Aesar, 99.99\%) at 1200~$^{\circ}$C in air for 80~h. Co$_3$O$_4$ was obtained by thermal decomposition of Co(NO$_3$)$_2$*6H$_2$O at 700~$^{\circ}$C in an oxygen flow. In order to obtain fully oxidized Sr$_2$Co$_{0.5}$Ir$_{0.5}$O$_4$ samples for spectroscopic studies, post-annealing in steel autoclaves at 400~$^{\circ}$C and 5000~bar O$_2$ pressure was performed for five days. The phase analysis and the determination of the unit cell parameters were performed using x-ray powder diffraction~\cite{Mikhailova17}. Transition metal cations Co and Ir are in edge-sharing oxygen octahedra, that are elongated along the c-axis. Reference compound NdCaCoO$_4$ was prepared by a solid state reaction. Starting materials of Nd$_2$O$_3$, CaCO$_3$ and Co$_3$O$_4$ were mixed in a stoichiometric ratio and ground thoroughly in an agate mortar, and then sintered in air at 1150 $^{\circ}$C for about 7 days with several intermediate grindings. The sample was post-annealed in 5000 bar O$_2$ at 450 $^{\circ}$C for 1 day.

Co-$L_{3,2}$ x-ray absorption spectra (XAS) of Sr$_2$Co$_{0.5}$Ir$_{0.5}$O$_4$ were recorded at the 08 beamline of the National Synchrotron Radiation Research Center (NSRRC) in Taiwan. CoO and NdCaCoO$_4$ samples were also measured as Co$^{2+}$ and Co$^{3+}$ reference compounds, respectively. All Co-$L_{3,2}$ spectra were collected at room temperature using total electron yield mode (by measuring the sample drain current) with an energy resolution of about 0.25 eV.

XMCD spectra at the Co-$L_{3,2}$ edges of Sr$_2$Co$_{0.5}$Ir$_{0.5}$O$_4$ were collected at the DEIMOS beamline \cite{Ohresser14} of SOLEIL in Paris (France) with a photon-energy resolution of 0.4 eV and a degree of circular polarization close to 100\%. The sample was measured at $T$~=~20~K and in a magnetic field of 6 Tesla. The spectra were recorded using the total electron yield method. The sample was cleaved $in~situ$ in order to obtain a clean surface for total electron yield measurements. The pressure during the measurements was below 5*10$^{-10}$~mbar.

\begin{figure}[ht]\centering
\includegraphics[width=\linewidth]{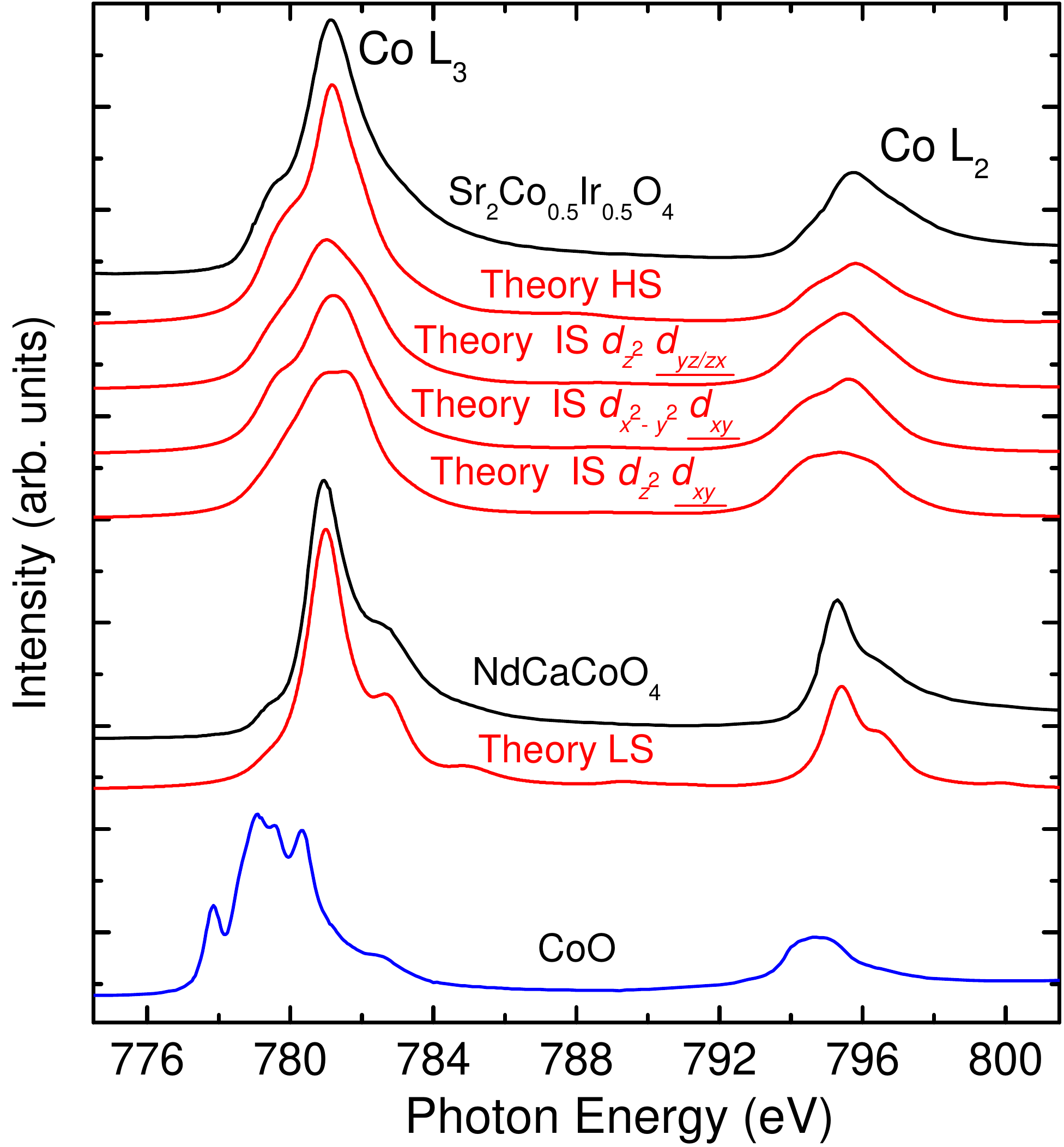}
\caption{Experimental (black lines) and calculated (red lines) Co-$L_{2,3}$ absorption spectra of Sr$_2$Co$_{0.5}$Ir$_{0.5}$O$_4$ together with NdCaCoO$_4$ as a LS-Co$^{3+}$ reference, and CoO as a Co$^{2+}$ reference (blue line). }\vspace{-0.2cm}
\end{figure}

In order to evaluate directly the local electronic structure of the Co ion in Sr$_2$Co$_{0.5}$Ir$_{0.5}$O$_4$, we also collected the Co-$L_{3,2}$ spectrum of NdCaCoO$_4$ as a reference compound for LS Co$^{3+}$ with the same K$_2$NiF$_4$ structure\cite{Li16}, and of CoO as a reference for HS Co$^{2+}$, see Fig. 1. The "center of gravity" of the $L_3$ white line of Sr$_2$Co$_{0.5}$Ir$_{0.5}$O$_4$ lies about 2~eV higher in energy than that of CoO, and only 0.1~eV higher than that of NdCaCoO$_4$, which confirms the Co$^{3+}$ valence in the Ir-compound. An Ir$^{5+}$/Co$^{3+}$ scenario was already suggested by our previous hard XAS study \cite{Mikhailova17}, although the Co $K$-edge XAS used there cannot give accurate Co valence estimation and is not very sensitive to the presence of Co$^{2+}$ impurities. The lack of a shoulder in the Co-$L_{3,2}$ spectrum at 777.8~eV, which is a fingerprint of HS Co$^{2+}$ with an octahedral local symmetry, clearly indicates the absence of any Co$^{2+}$ impurities in the Sr$_2$Co$_{0.5}$Ir$_{0.5}$O$_4$ sample.

Having ascertained the purity of the sample, we now focus on the determination of the Co$^{3+}$ spin state. In this context, it is important to note that the multiplet structure of the x-ray absorption spectrum depends strongly on the valence, the orbital and spin state. Hence, the spectral line shape can be used as fingerprint to determine the spin state. Despite having the same Co$^{3+}$ valence, the line shape of the Co $L$-edge spectrum of Sr$_2$Co$_{0.5}$Ir$_{0.5}$O$_4$ is very different from that of NdCaCoO$_4$, which suggests the two materials have different spin states. In order to understand this difference in spectral features we have performed a quantitative analysis of the spectrum by using the well proven configuration interaction cluster model that includes the full atomic multiplet theory \cite{Haverkort06}. The calculations were performed using the XTLS code\cite{Tanaka94}. We used a set of parameters close to that of LaCoO$_3$~\cite{Haverkort06}, but now considering also the tetragonal distortion effect on the crystal field \cite{calc_par}. The Co $3d-$O 2$p$ hybridization strengths were estimated according to Harrison's prescription\cite{Harrison}. The resulting energy level diagram is given in Fig. 2: NdCaCoO$_4$ in the left (a) part of the figure and Sr$_2$Co$_{0.5}$Ir$_{0.5}$O$_4$ in the right (b) part. The spin state of the energy levels is indicated by the colours: LS (red), IS (green), HS state (blue), and a mixture of LS/IS (orange). The calculated x-ray absorption spectra are plotted in Fig. 1 as red curves.

\begin{figure}[ht]\centering
\includegraphics[width=\linewidth]{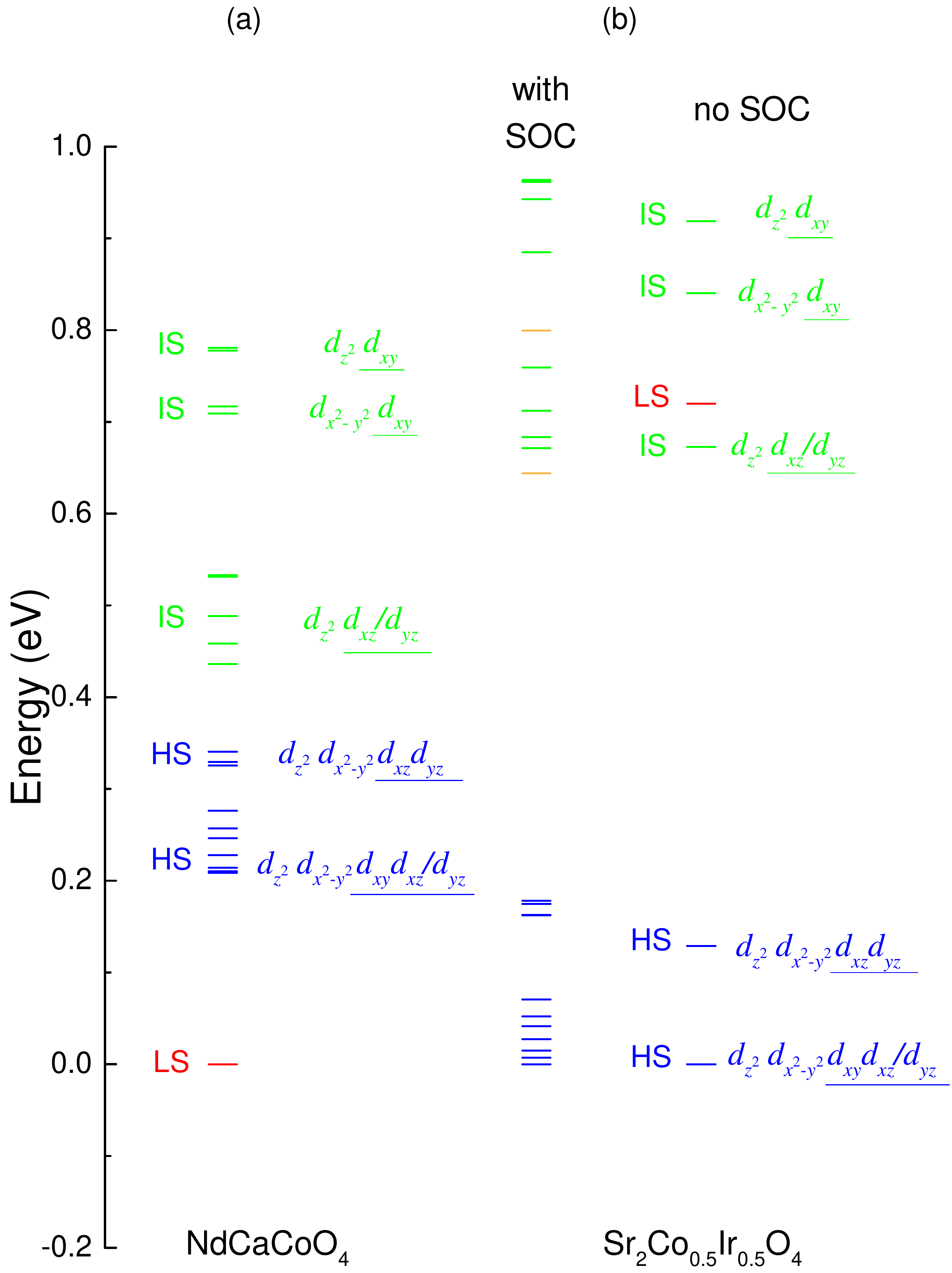}
\\
\caption{Energy level diagram of Co$^{3+}$ in NdCaCoO$_4$ (a) and Sr$_2$Co$_{0.5}$Ir$_{0.5}$O$_4$ (b). The spin state of the energy levels is indicated by the colours: LS (red), IS (green), HS state (blue), and a mixture of LS/IS (orange). The energy level diagram for the Sr$_2$Co$_{0.5}$Ir$_{0.5}$O$_4$ case is also presented with the spin-orbit coupling (SOC) switched off.}\vspace{-0.2cm}
\end{figure}

The energy level diagram of NdCaCoO$_4$ has the LS state as the lowest state, see Fig. 2(a), and the corresponding calculated Co-$L_{2,3}$ spectrum is displayed in Fig. 1 (bottom red curve). We can clearly observe that the experimental spectrum of NdCaCoO$_4$ is nicely reproduced: all important features have very similar energy positions and intensities. For Sr$_2$Co$_{0.5}$Ir$_{0.5}$O$_4$ we have instead the HS state as the ground state, see Fig. 2(b). We have calculated and plotted the corresponding spectrum in Fig. 1 (top red curve). Also here we can find an excellent match between the calculated and measured spectrum for Sr$_2$Co$_{0.5}$Ir$_{0.5}$O$_4$. We can therefore conclude that the Co ion in NdCaCoO$_4$ is in the LS configuration while the Co ion in Sr$_2$Co$_{0.5}$Ir$_{0.5}$O$_4$ is in the HS configuration. We would like to note that the measured and calculated spectra of NdCaCoO$_4$ are very different from those of Sr$_2$Co$_{0.5}$Ir$_{0.5}$O$_4$, and yet, for each compound we can find an excellent match between experiment and simulation. This can be taken as an indication that this spin state assignment is quite robust.

We will now look at the IS scenario and investigate whether we can exclude it for Sr$_2$Co$_{0.5}$Ir$_{0.5}$O$_4$ on the basis of the spectra collected. As a start, we look at Fig. 2(b) and we can see that there are many different IS states possible. At the same time, we can notice that the IS states are very much higher in energy (by more than 0.5 eV) than the HS states. This means that the local distortions are by far not large enough to stabilize the IS state in the real crystal structure of Sr$_2$Co$_{0.5}$Ir$_{0.5}$O$_4$. If the IS state were to be stabilized, then it should be done by band formation as mentioned above \cite{Korotin96,Ou16}. We then infer that in such a case, the band width must be so large that it overcomes the 0.5 eV energy difference, thereby also making the relatively small Co $3d$ spin-orbit interaction energy no longer to be an important quantity. We therefore propose to carry out the calculations for the IS states with the spin-orbit coupling (SOC) switched off. The corresponding energy level diagram is also displayed in Fig. 2(b) and the calculated spectra are presented in Fig. 1 (middle red curves) for the three lowest IS states, namely of the type $d_{z^{2} }\underline{d_{xz}/d_{yz}}$, $d_{x^{2}-y^{2}}\underline{d_{xy}}$, and $d_{z^{2} }\underline{d_{xy}}$ in ascending energy order. Here, the underline denotes a hole, the "/" symbol indicates a hole or an electron shared by two orbitals, and $z^{2}$ is an abbreviation for $3z^2-r^2$. The notation is the same as that employed by Haverkort $et~ al.$\cite{Haverkort06} and considers the full $t_{2g}$ shell of LS as a starting point from which the other orbital configurations are generated. We would like to note that the IS state studied in \cite{Merz11} is quite different because the authors used there an extremely large $e_g$ splitting that does not apply to the material studied here.

Comparing the measured Sr$_2$Co$_{0.5}$Ir$_{0.5}$O$_4$ spectrum to the three IS simulations we can quickly reject the $d_{z^{2} }\underline{d_{xz}/d_{yz}}$ and $d_{z^{2} }\underline{d_{xy}}$ scenarios: the experimental features at the $L_3$ edge are quite dissimilar from that in the $d_{z^{2} }\underline{d_{xz}/d_{yz}}$ simulation, while the spectral structures at both the $L_2$ and $L_3$ edges of the experiment are very different from those of the $d_{z^{2} }\underline{d_{xy}}$ simulation. Only the $d_{x^{2}-y^{2}}\underline{d_{xy}}$ simulation seems to reproduce the measured spectrum quite well. It is interesting to note that this $d_{x^{2}-y^{2}}\underline{d_{xy}}$ type of state is in fact the IS state that was proposed in the original work of Korotin $\it{et~al.}$ \cite{Korotin96}. In order to resolve whether the Sr$_2$Co$_{0.5}$Ir$_{0.5}$O$_4$ has really the HS Co$^{3+}$ state or rather the $d_{x^{2}-y^{2}}\underline{d_{xy}}$ IS state, we can make use of an important characteristic of this particular Korotin IS state, namely that it is a real-space orbital state and thus will not carry any orbital momentum, while the HS state, instead, will have a large orbital moment\cite{Haverkort06}. XMCD is a powerful technique to investigate this.

\begin{figure}[ht]\centering
\includegraphics[width=\linewidth]{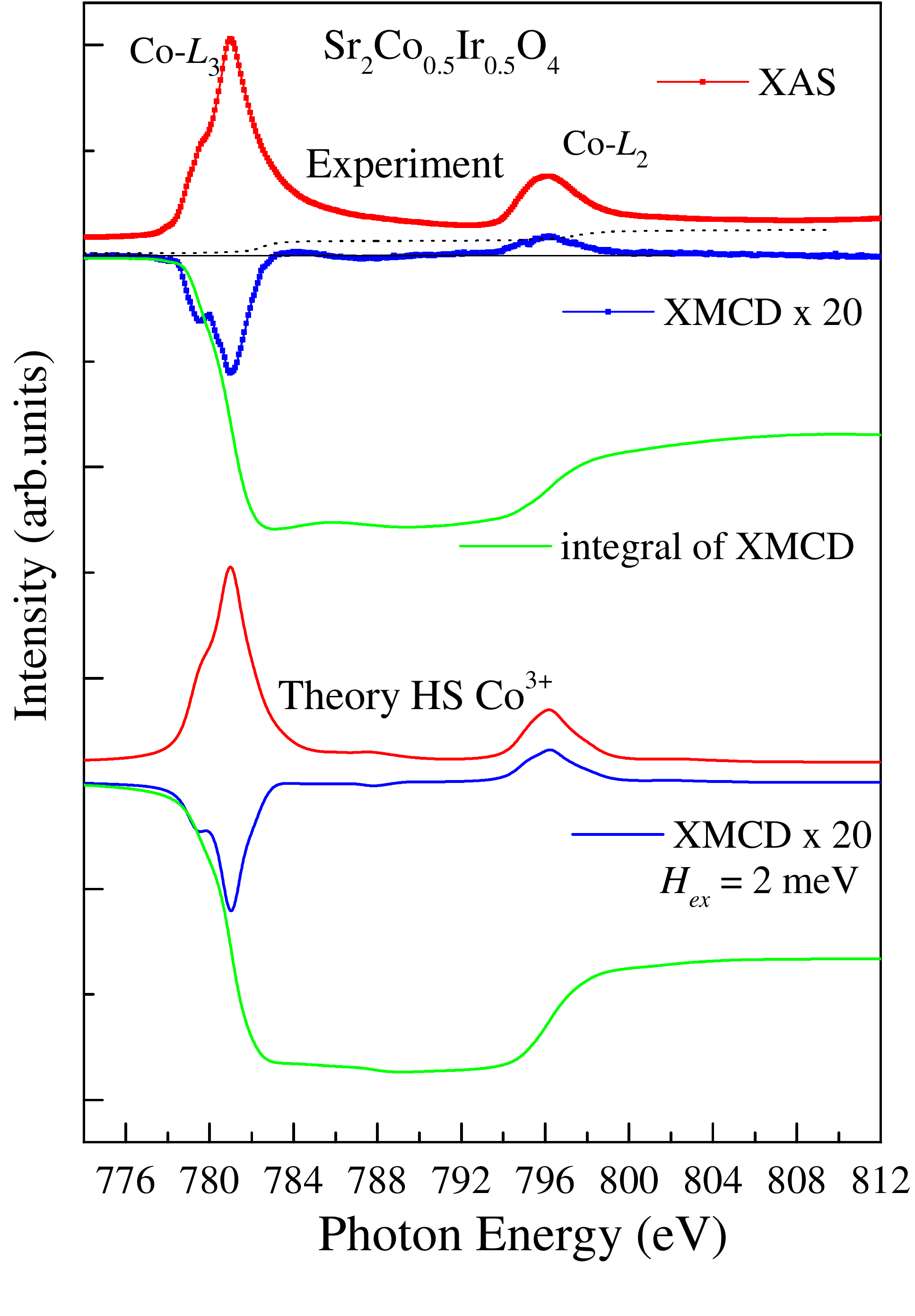}
\caption{Co-$L_{2,3}$ XAS (red squares) and XMCD  (blue squares) spectra measured on Sr$_2$Co$_{0.5}$Ir$_{0.5}$O$_4$ (top) compared with the calculated spectra (red and blue lines) for Co$^{3+}$ in a HS state (bottom). The green line indicates the integration over energy of the XMCD signal and the dotted line indicates the edge jump. }\vspace{-0.2cm}
\end{figure}

We have performed XMCD measurements at the Co-$L_{3,2}$ on Sr$_2$Co$_{0.5}$Ir$_{0.5}$O$_4$ and presented the results in Fig. 3. The x-ray absorption spectra were taken using circular polarized light with the photon spin parallel and antiparallel aligned to the magnetic field. The difference spectrum, called XMCD, and the sum spectrum, called XAS, are reported as blue and red curves, respectively. In Fig. 3 we have displayed also the theoretical Co-$L_{3,2}$ XAS and XMCD spectra for the Co$^{3+}$ in the HS configuration as obtained from our full-multiplet configuration-interaction calculations. Since we are dealing with a polycrystalline sample, we simulated the experimental data by summing two calculated spectra: one for light with the Poynting vector in the xy plane and one with the Poynting vector along the z-axis, with a weighting ratio 2:1. We can observe that the measured XMCD spectrum can be excellently reproduced by the HS simulation. This provides further evidence for the HS state of the Co ions in Sr$_2$Co$_{0.5}$Ir$_{0.5}$O$_4$. We would like to note that the XMCD spectrum of HS Co$^{3+}$ in octahedral symmetry has been rarely reported in literature: only in LaCoO$_3$ single crystal \cite{Haverkort06} and thin films \cite{Merz10,Mehta09}.

The large difference in intensity of the measured dichroic signal between the $L_3$ and $L_2$ edges shown in Fig. 3 is a clear sign that the Co ions have a significant unquenched orbital moment \cite{Thole92}. This then establishes directly that the Co ion in Sr$_2$Co$_{0.5}$Ir$_{0.5}$O$_4$ cannot be in the real-space IS state as discussed above. Instead we have a HS state that does carry orbital momentum despite the tetragonal distortion. To be quantitative, we include in Fig. 3 the integral of the XMCD signal over energy (green lines) of the measured and simulated spectra. We can clearly observe that the integrals converge to a finite non-zero value. We now apply the sum rules for XMCD developed by Thole $et~al.$ \cite{Thole92} and Carra $et~al.$ \cite{Carra93}, which provide the ratio:

\begin{equation}\label{eq:lz}
\frac{L_z}{2S_z+7T_z}=\frac{2}{3}\cdot \frac{\int_{L_{2,3}}(\sigma^+-\sigma^-)dE}{\int_{L_{3}}(\sigma^+-\sigma^-)dE-2\int_{L_{2}}(\sigma^+-\sigma^-)dE} ,
\end{equation}

\noindent
where $T_z$ denotes the intra-atomic magnetic dipole moment. For 3$d$ transition metal ions in an octahedral symmetry this term $T_z$ is a small number \cite{Teramura96} and is expected to be a little increased by the slight tetragonal distortion existing in the present compound. Indeed, from our configuration-interaction full-multiplet calculations we found the magnetic dipole moment for Co$^{3+}$ to be relatively small compared to the large spin moment of Co$^{3+}$ HS: $T_x/S_x =T_z/S_z = 0.05$. In other words, the important quantum number of $L_z/2S_z$ can be obtained directly from the experimental Co $L_{3,2}$ XMCD spectrum without requirement of theoretical simulations. From the application of the sum rules we obtained the ratio $L_z/2S_z$ = 0.25~\cite{sum_rules}. This value is similar to that of thermally populated HS Co$^{3+}$ ions observed in LaCoO$_3$ \cite{Haverkort06}. The unquenched orbital moment of the HS Co$^{3+}$ ions in Sr$_2$Co$_{0.5}$Ir$_{0.5}$O$_4$ is significant and might be estimated as $m(orb) = 0.76~\mu_B$ if we take the Co$^{3+}$ spin moment $m(spin) = 3.06~\mu_B$, as calculated in a previous theoretical work \cite{Ou14}.

It is worthy to note that the calculated XMCD signal for the HS Co$^{3+}$ in an applied field of 6 Tesla is 4.5 times larger than the measured one. A reduced XMCD signal could be explained by the presence of an antiferromagnetic arrangement of the Co$^{3+}$ moments, where the signal is given only by the canting of the moments induced by the applied field. Neutron diffraction measurements on our sample did not reveal the presence of a long-range magnetic order till the lowest temperature\cite{Mikhailova17}. However, the negative Weiss constant ($W = -27$~K) reveals AFM interactions between the Co ions, and a rise in the magnetic susceptibility with a maximum at $T^*=23$~K suggests the possible onset of a short range antiferromagnetic order of the Co ions in Sr$_2$Co$_{0.5}$Ir$_{0.5}$O$_4$ below $T^*$ \cite{Mikhailova17}. Following the hypothesis of a short range antiferromagnetic order, we introduced in the Hamiltonian an exchange field applied along the c-axis (the easy axis for an elongation distortion of the CoO$_6$ octahedron) and obtained the XMCD (reported in Fig. 3) as the average of the spectra calculated for opposite directions of the exchange field. An exchange field of about 2~meV is needed in order to reproduce the size of the experimental XMCD spectrum.

Pure HS Co$^{3+}$ is known to exist in tetrahedral CoO$_4$ \cite{Hollmann09} and square pyramidal CoO$_5$ coordination \cite{Belik06,Hu04}. However, for Co$^{3+}$ in octahedral local symmetry a pure HS state is rarely documented \cite{Ehrenberg09}, while the LS state \cite{Chang09,Hu04,Burnus06} or a mix of HS and LS states are usually found \cite{Haverkort06,Hu12}. One might wonder why Co$^{3+}$ is in a pure HS state in Sr$_2$Co$_{0.5}$Ir$_{0.5}$O$_4$, while NdCaCoO$_4$ shows Co$^{3+}$ in pure LS state, despite having exactly the same 214 layered structure.  Previous cluster calculations showed that the ground state of Co$^{3+}$ is in either HS or LS depending on the bond length, and a crossover LS-HS occurs at a Co-O distance of about 1.93~{\AA}~\cite{Chen14}. In the NdCaCoO$_4$ reference material, the short average Co-O bond length of 1.91~{\AA} corresponds to the LS state of Co$^{3+}$. The quite long average Co-O bond length (1.97~{\AA} from EXAFS measurements\cite{Mikhailova17}) of Sr$_2$Co$_{0.5}$Ir$_{0.5}$O$_4$ puts this material well in the region where the Co$^{3+}$ HS state is stable. Sr$_2$CoRuO$_{6-d}$ with even a longer average Co-O distance of 1.98~{\AA}, also provides a rare example of octahedrally coordinated HS Co$^{3+}$ ions~\cite{Chen14}. Yet, this compound is not stoichiometric in oxygen as demonstrated by the presence (7.5\%) of Co$^{2+}$ impurities. One, therefore, might still question whether oxygen vacancies stabilize the HS state in Sr$_2$CoRuO$_{6-d}$ like in many other Co$^{3+}$ materials with oxygen vacancies \cite{Hu12,Li11}. The lack of Co$^{2+}$ impurities shown by our XAS data, on the other hand, confirms that in the case of Sr$_2$Co$_{0.5}$Ir$_{0.5}$O$_4$ the HS state is not induced by oxygen vacancies.

\begin{figure}[ht]\centering
\includegraphics[width=\linewidth]{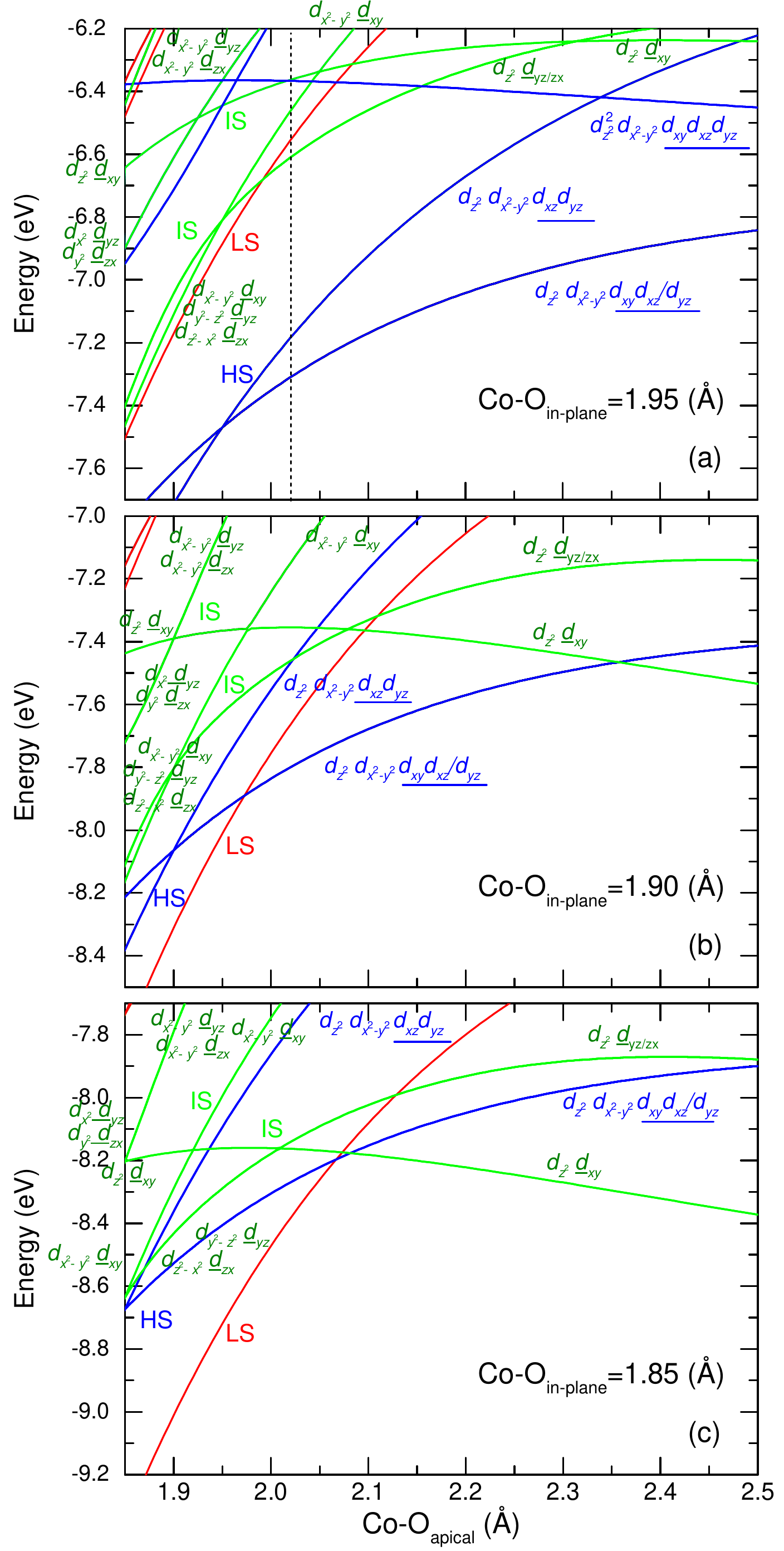}
\caption{Energy level diagram of Co$^{3+}$ as a function of the apical Co-O distance, for different in-plane Co-O distances : 1.95 {\AA} panel (a), 1.90 {\AA} panel (b), 1.85 {\AA} panel (c). The vertical dotted line indicates Sr$_2$Co$_{0.5}$Ir$_{0.5}$O$_4$. The spin state of the energy levels is indicated by the colours: LS (red), IS (green), and HS state (blue). The diagram is calculated with the SOC switched off.}\vspace{-0.2cm}
\end{figure}

The next question we would like to answer is what structural parameters might give rise to an IS ground state. To this aim we have calculated the energy level diagram of the Co$^{3+}$ ion (reported in Fig. 4) as a function of the apical Co-O distance for different values of the in-plane Co-O distance. In the calculations the ionic crystal field parameters were calculated using a point charge model, while the hybridization strengths were estimated according to Harrison's prescription\cite{Harrison}. We considered only the case of an elongated distortion, as a compressed distortion would induce a HS, not an IS state. Fig. 4(a) shows that for a relatively long in-plane Co-O distance, like 1.95~{\AA}, as in Sr$_2$Co$_{0.5}$Ir$_{0.5}$O$_4$, the HS is always the ground state no matter how large is the tetragonal elongation of the CoO$_6$ octahedra. In fact the effective energy splitting between the $t_{2g}$ and $e_g$ orbitals is too small to win against Hund exchange interaction, which favours a single occupation of the orbitals with parallel spin alignment. If the in-plane Co-O distance is reduced to 1.90~{\AA}, then the $d_{z^{2} }\underline{d_{xy}}$ IS state can become the lowest level in energy, as shown in Fig.4(b), but only when the apical Co-O distance is longer than 2.35~{\AA}. The $d_{z^{2} }\underline{d_{xy}}$ state is a real-space IS state and, thus, the orbital moment is zero. The apical distance required for stabilizing the $d_{z^{2} }\underline{d_{xy}}$ IS state is reduced if the in-plane distance is further decreased to 1.85~{\AA}, as shown in Fig.4(c). Therefore, to stabilize an IS state, an elongated Jahn-Teller distortion is not enough, but also a short in-plane Co-O distance (shorter than 1.91~{\AA}) is needed. At low temperature one of the two Co$^{3+}$ sites of the TlSr$_2$CoO$_5$ compound has the Co-O bond lengths Co-O$_{in-plane}=1.79$~{\AA} and Co-O$_{apical}=2.19$~{\AA} \cite{Doumerc99,Doumerc01}. It would be very interesting to perform an XMCD investigation of this compound to see whether the IS state is stabilized for one of the two Co$^{3+}$ sites.

To summarize, soft XAS and XMCD measurements at the Co-$L_{3,2}$ edge reveals a pure Co$^{3+}$ with high spin state in the hybrid 3$d$-5$d$ solid-state oxide Sr$_2$Co$_{0.5}$Ir$_{0.5}$O$_4$ with a layered K$_2$NiF$_4$ structure type. We attribute the stability of the Co$^{3+}$ HS state to the long average Co-O bond length in Sr$_2$Co$_{0.5}$Ir$_{0.5}$O$_4$. Our calculations predict that besides an elongated Jahn-Teller distortion, a short in-plane Co-O distance of less than 1.91~{\AA} is needed in order to have the Co$^{3+}$ in an IS ground state. The application of sum rules on the Co-$L_{3,2}$ XMCD data of Sr$_2$Co$_{0.5}$Ir$_{0.5}$O$_4$ gives nearly the same $L_z/2S_z$ ratio of about 0.25 as observed in LaCoO$_3$. Our XMCD study demonstrates, thus, the existence of a significant unquenched orbital moment of the Co$^{3+}$ ions despite the CoO$_6$ octahedra are sizeable elongated in Sr$_2$Co$_{0.5}$Ir$_{0.5}$O$_4$.

The research in Dresden was partially supported by the Deutsche Forschungsgemeinschaft through SFB 1143 and by the BMBF, project grant number 03SF0477B (DESIREE). K. Chen benefited from support of the German funding agency DFG (Project  600575).


\begin{thebibliography}{40}
\bibitem{calc_par} CoO6 cluster parameters (in eV): Slater integrals 80\% of the Hartree Fock value, spin-orbit coupling  $\zeta_{3d}$= 0.074, $Udd=5.5$, $Upd=7.0$, charge transfer $\Delta_{CT}=2$. For HS: 10$Dq_{ionic}=0.34$, $\Delta t_{2g}=0.08$, $\Delta e_g=0.14$, hybridization $V(b_{1g}) = 2.79$, $V(a_{1g}) = 2.55$, $V(b_{2g}) = 1.48$, $V(e_g) = 1.37$, ligand-crystal field 10$Dq_{ligand}=1$. Taking in account the effect of the hybridization with the oxygens, the effective splittings of the $3d$ orbitals are 10$Dq^{eff}=1.06$,  $\Delta t_{2g}^{eff}=0.13$,  $\Delta e_g^{eff}=0.31$. For LS: 10$Dq_{ionic}$=0.87,  $\Delta t_{2g}= \Delta e_g=0.0$, hybridization $V(b_{1g}) = 2.88$, $V(a_{1g}) = 2.50$, $V(b_{2g}) = 1.54$, $V(e_g) = 1.38$, ligand-crystal field 10$Dq_{ligand} =1$, which give the effective splittings of the $3d$ orbitals 10$Dq^{eff}=1.57$,  $\Delta t_{2g}^{eff}=0.1$,  $\Delta e_g^{eff}=0.31$. For all simulations reported in Fig. 1 and Fig. 3 we have considered the thermal population of the different states using a Boltzmann distribution.
\bibitem{sum_rules} From the separate application of the spin and orbital sum rules to our XMCD data we obtained $2S_z = 0.20 ~\mu_B$ and $L_z = 0.05~\mu_B$. These values are in very good agreement with that given by our simulations ($2S_z = 0.22~\mu_B$ and $L_z = 0.05~\mu_B$) reported in Fig. 3.
\end{thebibliography}
\end{document}